\documentclass[manuscript]{aastex631}

\usepackage{amsmath}
\usepackage{bm}

\def\vec#1{\ensuremath{\bm{#1}}}

\def \arcsec {$^{''}$}

\def \kms {{\rm km\;s$^{-1}$}}

\def\typei{Type-\uppercase\expandafter{\romannumeral1}~}
\def \typeii{Type-\uppercase\expandafter{\romannumeral2}~}
\def \paperi{Paper~\uppercase\expandafter{\romannumeral1}~}
\received{}
\revised{}
\accepted{}

\shorttitle{Kink Waves in Jets and Plumes}
\shortauthors{Qi et al.}

\begin{document}

\title{Propagating Kink Waves in Chromospheric Jet-like Structures and Coronal Plumelets}

\correspondingauthor{Lidong Xia, Mingzhe Guo}
\email{xld@sdu.edu.cn, m.guo@sdu.edu.cn}

\author{Youqian Qi}
\affiliation{Shandong Key Laboratory of Space Environment and Exploration Technology, Institute of Space Sciences, Shandong University, Shandong, China}

\author{Mingzhe Guo}
\affiliation{Institute of Frontier and Interdisciplinary Science, Shandong University, Qingdao, 266237, China}
\affiliation{Shandong Key Laboratory of Space Environment and Exploration Technology, Institute of Space Sciences, Shandong University, Shandong, China}

\author{Zhenghua Huang}
\affiliation{Institute of Science and Technology for Deep Space Exploration, Suzhou Campus, Nanjing University, Suzhou 215163, People's Republic of China}

\author{Tom Van Doorsselaere}
\affiliation{Centre for mathematical Plasma Astrophysics (CmPA), Department of Mathematics, KU Leuven, Celestijnenlaan 200B, 3001 Leuven, Belgium}

\author{Bo Li}
\affiliation{Shandong Key Laboratory of Space Environment and Exploration Technology, Institute of Space Sciences, Shandong University, Shandong, China}

\author{Lidong Xia}
\affiliation{Shandong Key Laboratory of Space Environment and Exploration Technology, Institute of Space Sciences, Shandong University, Shandong, China}

\author{Hengyuan Wei}
\affiliation{Institute of Science and Technology for Deep Space Exploration, Suzhou Campus, Nanjing University, Suzhou 215163, People's Republic of China}

\author{Hui Fu}
\affiliation{Shandong Key Laboratory of Space Environment and Exploration Technology, Institute of Space Sciences, Shandong University, Shandong, China}

\begin{abstract}
Coronal plumes and chromospheric jet-like structures are believed to be highly dynamic. 
We report the first direct observations of a propagating kink wave in a chromospheric jet-like structure 
and its associated plumelet structure in the upper corona of the solar polar region, 
using data from the High Resolution Imager (HRI) of the Extreme Ultraviolet Imager (EUI) on board Solar Orbiter (SO).
The dark jet-like structure exhibits transverse oscillation during upward propagation, with a period of approximately 
95\,s and a displacement of about 193\,km. The corresponding plumelet also displays transverse motion, 
with an oscillation period of around 99\,s and a displacement of about 315\,km. 
Given that both the dark jet-like strucutre and the plumelet share the same magnetic skeleton and have similar oscillation period, 
we suggest that these oscillations are the same transverse propagating wave originating in the chromosphere.
This scenario is further supported by a 3D magnetohydrodynamic (MHD) simulation, 
in which both vertical and transverse perturbations were introduced in a stratified magnetic flux tube. 
The simulation successfully reproduces the upward propagation of a kink wave through both the chromospheric jet-like structure and the coronal plumelet. 
These results highlight the potential role of transverse waves in transferring energy from the lower solar atmosphere to the corona.

\end{abstract}

\section{INTRODUCTION} 
\label{sec_intro}
Coronal plumes belong to bright and collimated structures off the solar limb, which are relatively large and stable~\citep[see][and the references therein]{2011A&ARv..19...35W}. 
They have a base diameter of 20 to 30\,Mm and lifetimes ranging from $\sim$20 hours to days. 
Some works have indicated that coronal plumes might feed sufficient plasma and energy into the fast solar wind~\citep{2010ApJ...709L..88T,2011ApJ...736..130T,2015SoPh..290.1399F}.
With high spatio-temporal resolution observations,
\citet{2021ApJ...907....1U} revealed the presence of dynamic filamentary structures within plumes, termed plumelets.
These features are likely the primary energy sources for coronal plumes~\citep[see][and the references therein]{2008ApJ...682L.137R,2014ApJ...787..118R,2016SSRv..201....1R}.
\citet{2022ApJ...933...21K} proposed that these transient jetlets, which result from reconnection at the open-closed boundary, can carry mass and energy, extending upward to form plumelets. 
These plumelets exhibit transverse widths of 10\,Mm and intermittently support upwardly propagating disturbances with speeds ranging from 190 to 260\,\kms. 

Plumelets generally exhibit upward and transverse motions.
Quasi-periodic propagating disturbances (PDs) are typical vertical motions in plume 
and are characterized by periods of 10 to 30 minutes and propagation speeds of 75 to 300\,\kms~
\citep[e.g.][]{1997SoPh..175..393D,1998ApJ...501L.217D,2000SoPh..196...63B,2009A&A...499L..29B,2011A&A...528L...4K}. 
PDs are attributed to slow density disturbances or outflows~\citep{1998ApJ...501L.217D,1998A&A...339..208B,2009A&A...501L..15B,2011ApJ...736..130T,2012ApJ...754..111O}. 
On the other hand, 
transverse motions in plumes are believed to be kink waves. \citet{2014ApJ...790L...2T} reported transverse displacements in a broad range in plumes via SDO/AIA observations. 
Furthermore, \citet{2018ApJ...852...57W} found that about 40\% of observed plumes exhibit transverse motions.
These transverse waves have periods of 55 -- 291\,s and amplitudes of 110 -- 704\,km.

Many studies have revealed a strong relationship between coronal plumes (or PDs within coronal plumes) and jet-like structures in the lower atmosphere 
\citep[e.g.,][]{2015ApJ...809L..17J,2019Sci...366..890S,2024A&A...689A.135S,2025ApJ...979..195D}.
In chromospheric spectral lines,
spicules are one type of the jet-like features off the solar limb \citep{1968SoPh....3..321N}. 
In TRACE and AIA observations,
\citet{2019SoPh..294...96A} proposed the absorption features identified off the limb in the 171 \AA ~are also associated with spicules. 
These features exhibit characteristics similar to those of spicules identified via emission features in the 1600 \AA.
Based on their trajectories of motion, spicules are generally classified into two types~\citep{2007PASJ...59S.655D}. 
\typei spicules are primarily excited by photospheric p-mode waves~\citep{2006ApJ...647L..73H, 2007ApJ...655..624D}. 
These spicules exhibit upward motion and then fall back to the solar surface, 
with typical speeds of $\sim$ 25\,\kms and lifetimes on the order of a few minutes.
\typeii spicules are primarily driven by magnetic reconnection~\citep{1995Natur.375...42Y,2018ApJ...852...16Y,2019Sci...366..890S}. 
These spicules go upward and then vanish in chromospheric spectral lines, with typical speeds of $\sim$ 100\,\kms and lifetimes of approximately 100\,s. 
Both types are highly dynamic and have fine-scale features.
Based on high-order solar adaptive optics,~\citet{2025NatAs...9.1148S} found fine and dark prominences off the solar limb in H$\alpha$ lines. 
 These prominences have the width of 150 - 300\,\kms and show twisted characters. 
 They exhibit turbulent flow, and go upward with approximately constant speeds of $\sim$ 20\,\kms \citep{2008ApJ...676L..89B}.

These jet-like structures usually exhibit swaying motions, 
which are considered indicative of kink waves propagate along their axis \citep{2008ApJ...676L..89B,2009A&A...497..525H,2007Sci...318.1574D,2011ApJ...736L..24O}. 
Such waves are capable of transporting sufficient energy to heat the plasma in transition region or even in the corona.
In \citet{2007Sci...318.1574D}, for example, 
the velocity amplitudes of the transverse motion range from 10 to 25\,\kms, the periods spanning from 100 to 500\,s. 
Furthermore,
\citet{2011ApJ...736L..24O} reported both standing and propagating transverse waves in spicules with the medians of the period and velocity amplitude of 45\,s and 7.4\,\kms.
\citet{2009A&A...497..525H} also reported high-frequency ($>$ 0.01\,Hz) transverse displacements in spicules, 
which might result from small-scale magnetic reconnection~\citep{1999SSRv...87...25A}. 
An alternative perspective holds that transverse waves can be excited by torsional motions in the chromosphere,
which originated from vortex motions of a strong magnetic flux concentration in the photosphere \citep{2013ApJ...768...17M}. 
In fact,
\citet{2014ApJ...788....9G} proposed that the displacement of kink waves consists of both transverse and rotational components.

In general, kink waves are one potential candidate for transferring energy from the
lower atmosphere into the corona.
\citet{2011Natur.475..477M} confirmed that outward propagating kink waves in spicules are energetic enough to heat the quiet corona.
However,
whether kink waves can propagate from the lower chromosphere to the corona or not still lack direct observational evidence. 
This paper aims to investigate the propagating kink waves from chromospheric jets to coronal plumelets,
from both high-resolution observational and numerical perspectives.
The observational data analysis and simulations are described in Sect.~\ref{sec:data} and Sect.~\ref{sec:simulation}, respectively. 
The findings are summarized and discussed in Sect.~\ref{sec:con}.

\section{Observations and data analysis}
\label{sec:data}

\begin{figure*}[!ht]
\centering
\includegraphics[trim=0cm 0cm 0cm 0cm,clip,width=\textwidth]{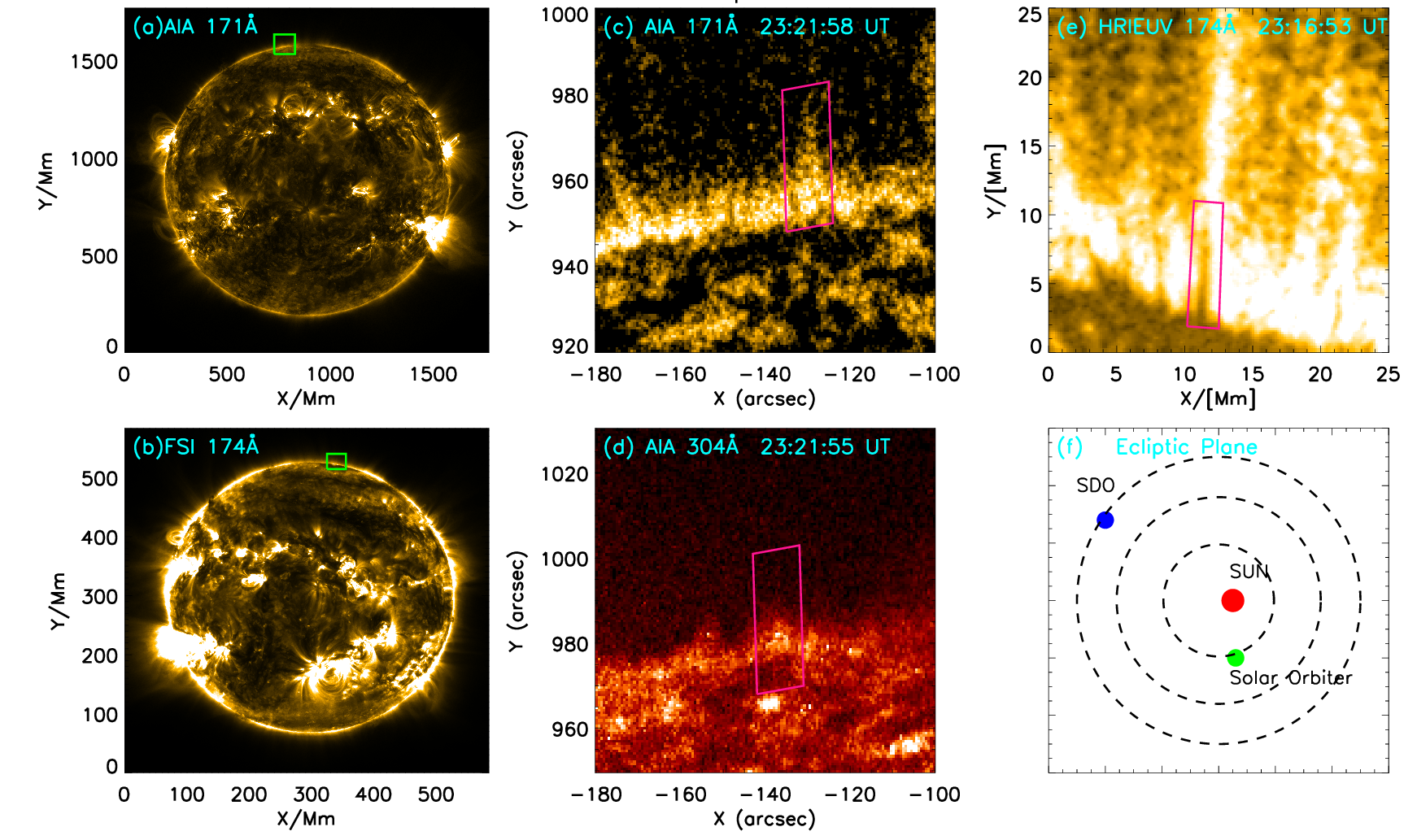}
\caption{Context images for structures studied in the present work taken on 24 April 2023.
The AIA 171\,\AA~and FSI/SO 174\,\AA~images are giving an overview in panel (a) and (b).
The region enclosed by the rectangle (green lines) in panel(a) is zoomed-in in AIA 171\,\AA (c) and 304\,\AA(d).
The region enclosed by the rectangle (green lines) in panel(b) is zoomed-in in EUI 174\,\AA (e).
The pink square mark our target structure from different views.
Panel (f) displace the  position of SDO and SO in Heliocentric Earth Ecliptic (HEE) coordinates system.
The black dash circles indicate the Line-of-Sight (LOS) of HRIEUV/SO and AIA/SDO.
}
\label{fig_orbit}
\end{figure*}

\begin{figure*}[!ht]
\centering
\includegraphics[trim=0cm 0cm 0cm 0cm,clip,width=\textwidth]{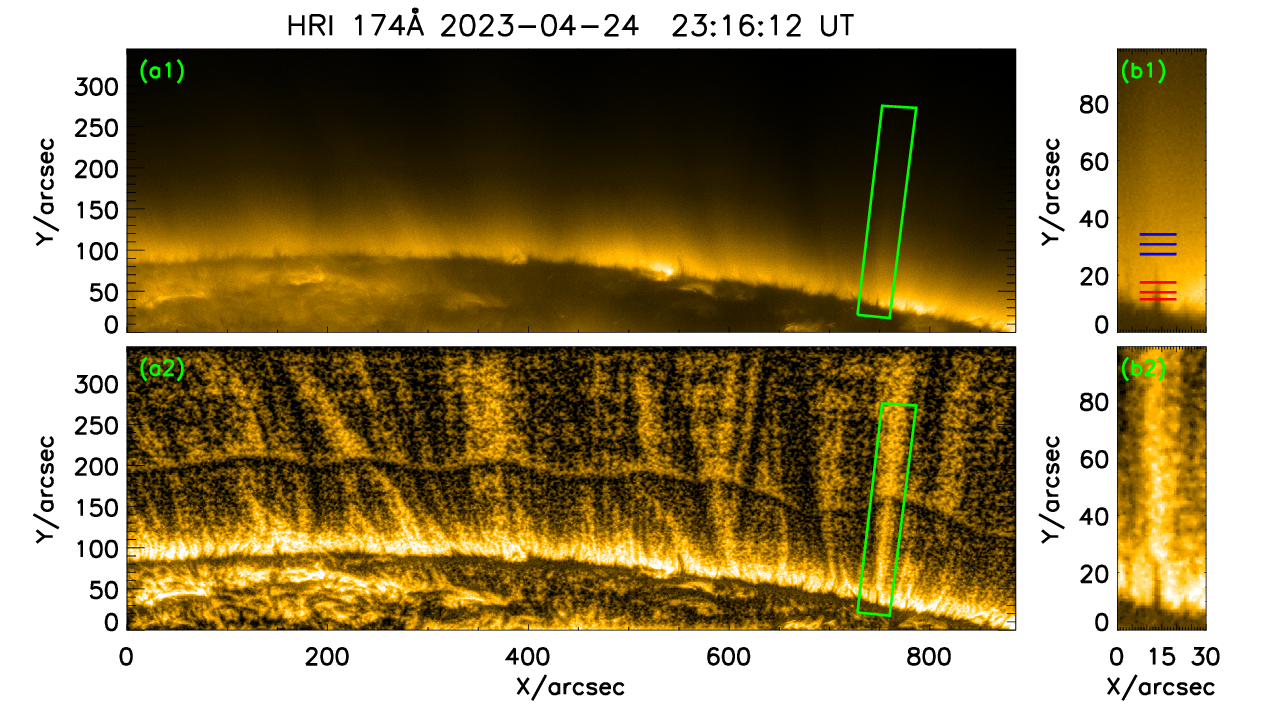}
\caption{The field of HRI$_{EUI}$ 174\,\AA~observations of darkjet-like structures and plumelet. 
Panel (a1) shows the original observations, and panel (a2) shows the optimized image using the WOW technique. 
Green rectangles show the concerned structures and are zoomed in on the right. Red (blue) lines in panel (b1) 
at $h=1.9\,Mm, 2.5$\,Mm, 3.4\,Mm ($h=6.5\,Mm, 7.6$\,Mm, 8.6\,Mm) marked the location of the corresponding slit across the observed structure. 
An animation of the density evolution is available on line, 
the amnimated figure runs from 23:15:00 UT to 23:44:44 UT, with a total duration of 1784 seconds.
}
\label{fig_fov}
\end{figure*}


\begin{figure*}[!ht]
\centering
\includegraphics[width=\textwidth]{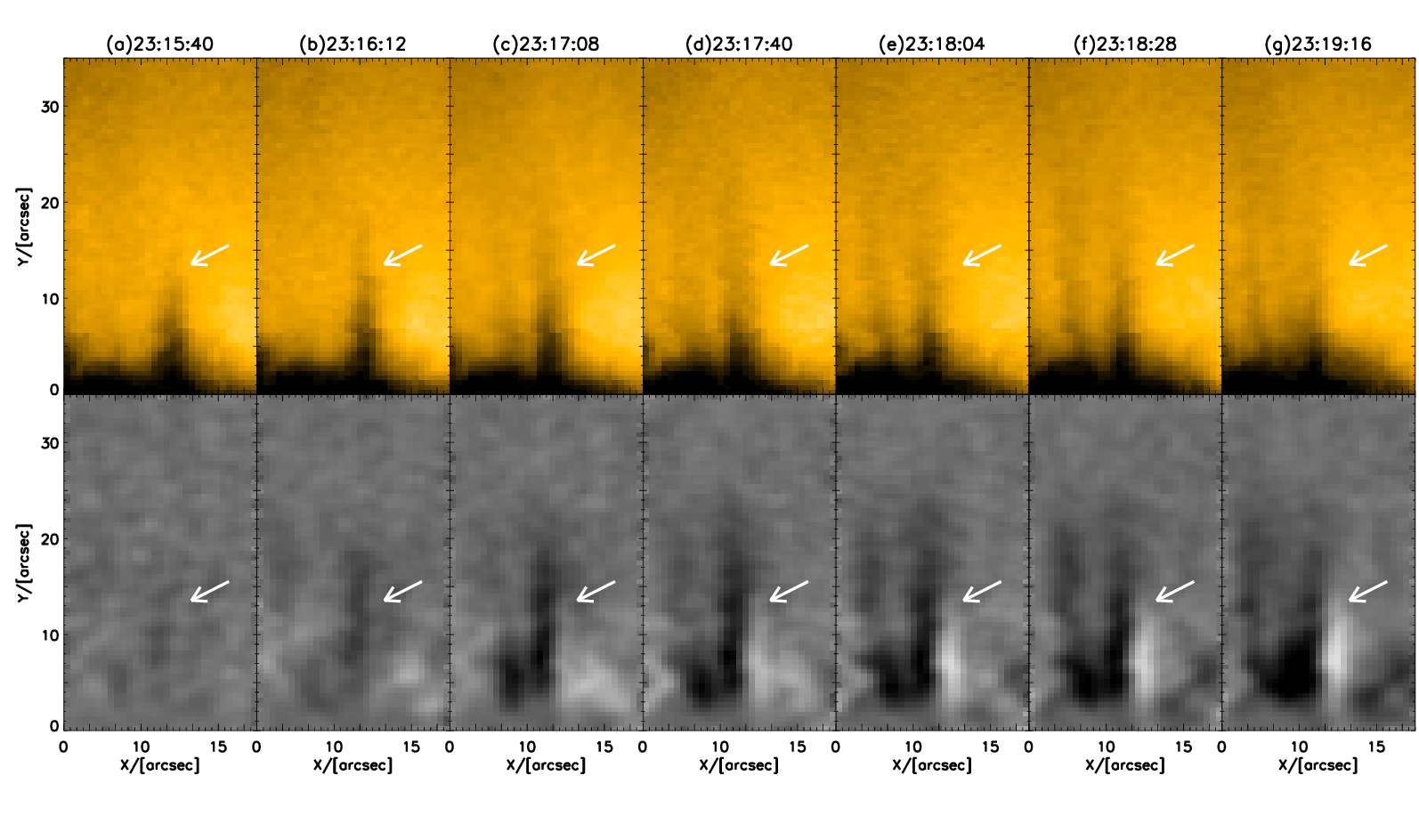}
\caption{The evolution of the chromospheric darkjet-like structures (marked as the white arrow) we analyzed. 
The upper row shows the original images, and the lower row shows the difference images.}
\label{jet_vertical}
\end{figure*}


The dataset analyzed in this study was acquired on 24 April 2023, from 23:15 UT to 23:44 UT, 
focusing on the solar polar region off the solar limb. 
Data were collected using the High Resolution Imager (HRI) of the Extreme Ultraviolet Imager
~\citep[EUI,][]{2020A&A...642A...8R} 174\,\AA~passband aboard the Solar Orbiter~\citep[SO,][]{2020A&A...642A...1M}.
The temporal and spatial resolution of HRI$_{EUI}$ 174\,\AA~is  8\,s and 0.492\arcsec (i.e., 146 km/pixel).
AIA 304\,\AA~took images of the same jet from a different lines-of-sight (LOS), where it appeared bright and fuzzy, see Figure~\ref{fig_orbit}(d). 
As the spatial resolution of AIA images ( 435\,km/pixel ) is about 3 times HRI$_{EUI}$'s, 
we employed coronal plume structures observed by the AIA 171\,\AA~channel for comparative analysis, 
in order to confirm the location of the jet detected in the 304\,\AA~channel.
During the observation, the distance from the Sun for SO is 0.42\,AU, and that for SDO is 1.0\,AU. 
\textbf{Therefore, the light travel time difference between two spacecrafts was 290\,s (4.8\,min).}
See their location in Figure~\ref{fig_orbit}, the angle separation between the two LOS was 132.3 degrees.
No distinct loop structures are observed at either viewing angle,  and the structure is more consistent with a jet interpretation.

Figure~\ref{fig_fov}(a1) shows an overview of the observed region.
Bright, ray-like features are identified as coronal plumes,
while dark, jet-like features are interpreted as chromospheric absorbing structures, 
as described in~\citet{2019SoPh..294...96A}. 
To improve the visibility of substructures (plumelets) within the plumes,
we applied the wavelet-optimized whitening (WOW) enhancement technique 
\citep{2023A&A...670A..66A}, as shown in Figure~\ref{fig_fov}(a2). 
This method works by equalizing the variance at all scales and locations in an image, 
thereby reducing the large gradients and, conversely, enhancing fine structures. 
The green rectangles in Figure~\ref{fig_fov}(a1 \& a2) delineate the region of interest, 
with zoomed-in images on the right.
In Figure~\ref{fig_fov}(b1), different slits are used to capture the intensity distribution across 
(red lines) the dark jet-like structures and 
across (blue lines) the coronal plumelet. 
The red lines from bottom to top represent the jet-like structure at heights of 1.9\,Mm, 2.5\,Mm and 3.4\,Mm, respectively. 
The blue lines from bottom to top represent the plumelet at heights of 6.5\,Mm, 7.6\,Mm and 8.6\,Mm, respectively.

\subsection{Dynamics of Chromospheric Jet-like Structures}\label{sec:dark jet}

\begin{figure*}[!ht]
\centering
\includegraphics[width=0.9\textwidth]{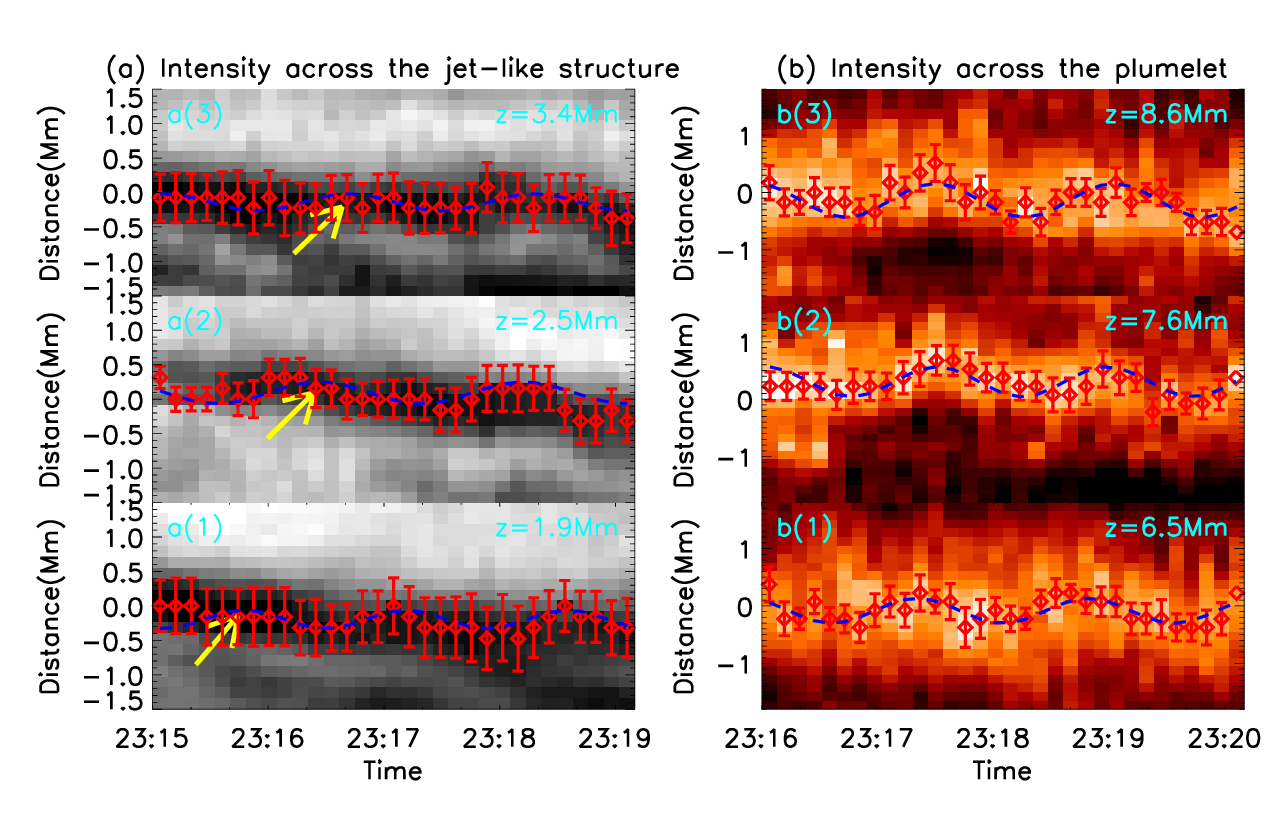}
\caption{Diagrams of the transverse displacement versus time for the 174\,\AA~ intensity 
across (a1-a3) the dark jet-like structure and (b1-b3) the plumelet. 
The centroid of the transverse profile is plotted as red diamonds. 
Vertical red bars show the 2$\sigma$ width. 
The blue dashed lines are the sine-fitting results according to Eq. \eqref{fit_eq1} and Eq. \eqref{fit_eq2}.}
\label{across_jet}
\end{figure*}

The chromospheric dark jet-like structure exhibits highly dynamic behavior. 
Analysis of the time-series animation reveals unambiguous evidence of 
both vertical and transverse motions associated with this structure.
Seen from the white arrow in Figure~\ref{jet_vertical},
the dark jet-like structure moves upward first and then drops down.
We can determine that 
the height of the structure can reach is 3.8\,Mm, propagating upward at an apparent velocity of approximately 26\,\kms.
These characteristics of the dark jet-like structures are similar to those of traditional spicules~\citep{2007PASJ...59S.655D,2008ApJ...676L..89B,2025NatAs...9.1148S}.
Furthermore, the similar parabolic trajectory of a chromospheric jet has been reported in previous studies of pressure-driven spicules \citep[e.g.,][]{2010A&A...519A...8M,2024SoPh..299...53Z}.
This probably indicates that the chromospheric dark jet-like structure discussed here is excited by pressure pulses or slow shocks. 
Regarding the transverse motion, 
Figure~\ref{across_jet}(a1 - a3) shows the time-distance map of intensity across the dark structures 
at  $h=1.9$\,Mm, $h=2.5$\,Mm and $h=3.4$\,Mm above the limb. 
The location of these slits are labelled by the red lines in Figure~\ref{fig_fov}(b1).
To improve the signal-to-noise ratio, we averaged the intensity data from two adjacent pixels along the
every red line. The central position of the jet-like structure corresponds to the  intensity-averaged extremum points, 
which is denoted by red diamonds.
Vertical red bars show the 2$\sigma$ width, 
with $\sigma$ being the square root of the variance for the \textbf{intensity-averaged} profile across the darkjet-like structure.
Similar procedure has been described in \citet{2009ApJ...705L.217H}.
During the lifetime of the jet from 23:15:48 to 23:19:00 UT, 
shown in Figure~\ref{across_jet}(a1 - a3), 
we can observe that the jet-like structure shows oscillations in the transverse direction. 
This kind of transverse motion has been reported in chromospheric spicules \citep[e.g.,][]{2011Natur.475..477M,2009ApJ...705L.217H}.
The oscillation profile in Figure~\ref{across_jet}(a) can be fitted by 
\begin{eqnarray}
D_1(t)=A_1\sin(2\pi t/P_1+\phi_1)+C_1+E_1\times t+F_1\times t^2,
\label{fit_eq1}
\end{eqnarray}
where $A_1$ is the displacement of the oscillation, $P_1$ represents the period of the oscillation, $\phi_1$ is the phase difference,
Besides, $C_1$, $E_1$ and $F_1$ are constant parameters determining the parabolic trend. 
These six parameters and their errors could be attained by MPFIT procedure from the SolarSoft (SSW) package, in which the $A_1$ and $P_1$ are most important ones.


\begin{figure*}[!ht]
\centering
\includegraphics[width=0.9\textwidth]{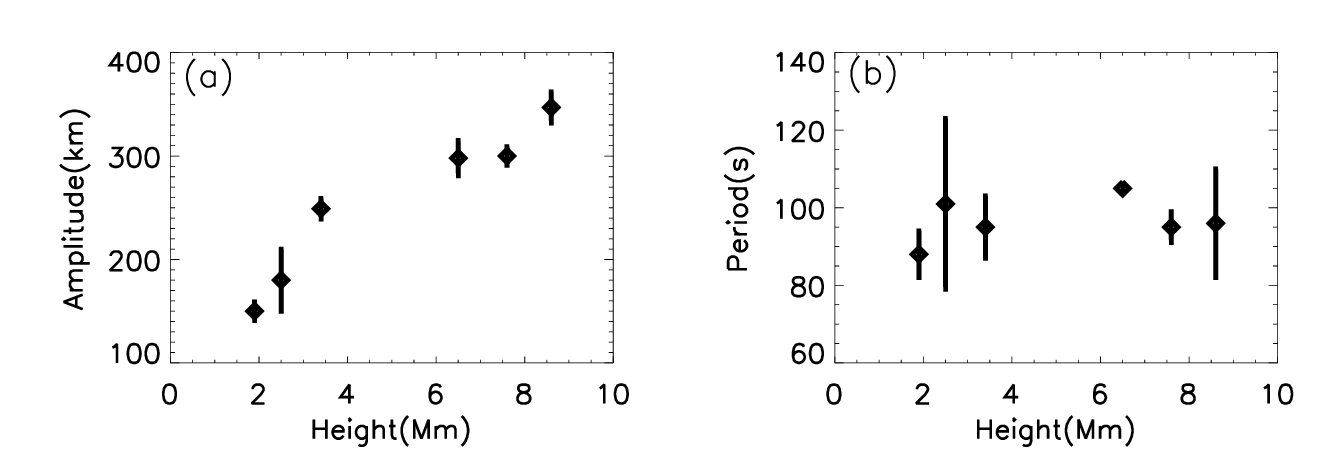}
\caption{\textbf{The parameters associated with the transverse motions of the jet-like structure and plumelet. 
Panel (a) displays the displacements and their errors of transverse oscillations at different heights, 
panel(b) displays the periods and their errors of transverse oscillations at different heights.}}
\label{tab_sin}
\end{figure*}


Figure \ref{across_jet}(a1 - a3) illustrates the further evolution of this transverse-displacement oscillation 
at higher sections of the jet-like structure trace. 
The blue dashed lines in the various panels of Figure \ref{across_jet}(a1 - a3) 
represent the sine-fitting results for the oscillation profiles obtained at different heights. 
The yellow arrows represent the propagation of the transverse-displacement oscillation of jet-like features. 
The amplitudes at heights of 1.9, 2.5 and 3.4\,Mm are 150 $\pm$ 9, 180 $\pm$ 30, 249 $\pm$ 10\,km, respectively, 
and the periods are 88 $\pm$ 6, 101 $\pm$ 22, 95 $\pm$ 8\,s, respectively.
It can clearly be seen that the displacement amplitude steadily increases up to the height, 
while the period remains constant at approximately 95 seconds from the Figure~\ref{across_jet} (a1 - a3) and \textbf{Figure~\ref{tab_sin}}.
Of particular note is that at a jet height of 1.9\,Mm, 
the measured oscillation amplitude becomes comparable to the instrumental resolution limit ($\sim$146\,km). 
Furthermore, the oscillatory signature of the jet-like features cannot be reliably resolved at altitudes below 1.9\,Mm. 
This observation provides evidence for the amplification of oscillation amplitude during its propagation.

\subsection{Transverse Oscillations in the Plumelet}\label{sec:plumelet}

Figure~\ref{across_jet}(b1 - b3) shows the time-distance map of the intensity across the plumelet at 
a height of 6.5\,Mm, 7.6\,Mm and 8.6\,Mm above the limb.
Similar to the dark jet-like structure shown in Figure~\ref{across_jet}(a1 - a3), 
the red diamonds represent the center of the plumelet structure, 
and the blue dashed lines give the fitting curve.
Similar to the procedure described in the previous section,
the oscillation curve of the plumelet is fitted by
\begin{eqnarray}
D_2(t)=A_2\sin(2\pi t/P_2+\phi_2)+C_2+E_1\times t+F_1\times t^2.
\label{fit_eq2}
\end{eqnarray}
Here,
$A_2$ represents the oscillation displacement of the plumelet, $P_2$ is the oscillation period of the plumelet, 
$\phi_2$ is the phase difference of the plumelet,
Besides, $C_2$, $E_2$ and $F_2$ are constant parameters determining the parabolic trend.

Figure \ref{across_jet}(b1 - b3) illustrates the further evolution of this 
transverse-displacement oscillation at higher sections of the plumelet trace. 
The blue dashed lines in the various panels of Figure \ref{across_jet}(b1 - b3) 
represent the sine-fitting results for the oscillation profiles obtained at different heights. 
The amplitudes at heights of 6.5, 7.6 and 8.6\,Mm are 298 $\pm$ 17, 300 $\pm$ 9, 347 $\pm$ 15\,km, respectively, 
and the periods are 105 $\pm$ 1, 95 $\pm$ 4, 96 $\pm$ 14\,s, respectively.
The displacement amplitude exhibits a steady increase with increasing height, 
a trend that is clearly discernible from the Figure~\ref{across_jet} (b1- b3) and Figure~\ref{tab_sin}.

Meanwhile, the oscillation displacement of the plumelet is larger than that of the dark jet-like structure from the Figure~\ref{tab_sin}. 
This is understandable because of the density stratification in the concerned structure \citep[see also similar discussion in][]{2024A&A...689A.195G}.
Note that the onset time of the plumelet oscillation is about 23:16:44\,UT, which is slightly delayed compared to that of the dark jet-like structures. 
Given also that the average oscillation period of the plumelet (99\,s) matches that of the dark jet-like structure,
we believe that the transverse wave observed in the plumelet originates from the chromospheric jet. 
Based on the vertical separation between the different heights, 
we derive a rough estimate of the transverse wave propagation velocity at approximately 91\,\kms. 
While this value appears relatively low, it is consistent with the Alfv\'en speed estimates reported in \cite{2007Sci...318.1574D}, 
thereby suggesting that the propagating waves investigated herein are likely kink waves.
Note that the propagating speed is roughly estimated,
as the onset time of the oscillation in the coronal counterpart is not straightforward to measure due to the resolution limit (see Figure~\ref{across_jet}).

Based on the observational evidence presented above, 
we conclude that the waves detected in the dark jet-like structure and plumelet 
correspond to kink waves propagating from the chromosphere to the corona. 
While the propagation velocity is comparatively low, 
implying a modest associated wave energy flux,
this constitutes the first direct observational detection of upward-propagating kink waves 
originating from the lower solar atmosphere and extending into the corona.

\section{Numerical Modelling}
\label{sec:simulation}

In the previous section, 
the transverse wave motion is a key property of the observed jet-like structure and plumelet. 
To clarify and confirm the dynamic relationship between the chromospheric jet and the coronal plume,
we conduct a numerical simulation to reproduce some key features of the above observations. 
Specially, the chromospheric jet and the plumelet share a same magnetic skeleton during the period we analyzed, 
as seen in Figure~\ref{fig_fov}(b2), 
we thus consider a pre-existing magnetic flux tube in the following simulation to mimic the magnetic structure we observed.

\subsection{Model Setup}

\begin{figure*}[!ht]
\centering
\includegraphics[width=\textwidth]{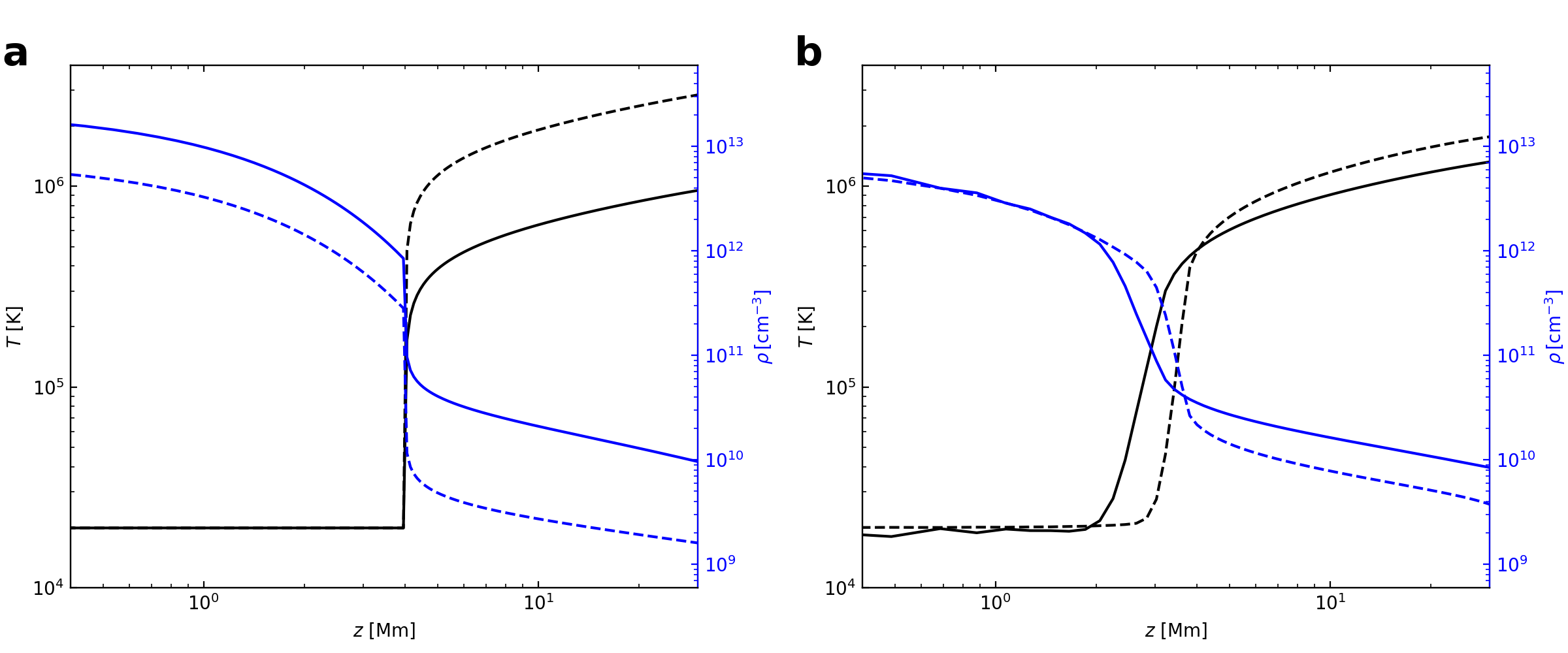}
\caption{Distributions of temperature (black curves) and density (blue curves) along the $z$-axis. Panel (a) shows the initial values before relaxation in the 2D run. 
Panel (b) displays the initial equilibrium of the 3D computation after relaxation. Solid and dashed lines represent the distribution at the loop axis ($x = 0, y = 0$) 
and the external loop region ($x = 6 {\rm Mm}, y = 0$), respectively.
}
\label{fig_ini_prof}
\end{figure*}

We consider a similar flux tube model employed in \cite{2023A&A...672A.105P} and \cite{2023ApJ...949L...1G}.
As described in \cite{2023ApJ...949L...1G},
the model is initiated from a 2D hydrostatic equilibrium.
Here, we just mention the difference from the previous model.
In cylindrical coordinates $(r, z)$,
the gravity is considered as
$g(z)=-g_\odot R_\odot^2/(z+R_\odot)^2$.
The initial uniform magnetic field is $\vec{B}=B_0\hat{z}$,
with $B_0$ being 20\,G.
The initial temperature profile is the same as in \cite{2023ApJ...949L...1G}.
The initial density and temperature are shown in Figure~\ref{fig_ini_prof}.
As the initial state is not in magnetohydrostatic balance,
a relaxation should be considered.
To maintain the stability of the simulation,
we consider a modification of the velocity as
$v'(r,z;t)=\alpha(t)\beta(z) v_{i}(r,z;t)$.
The expression of the temporal parameter $\alpha(t)$ is given by Eq. (3) in 
\cite{2023ApJ...949L...1G}.
In the present model,
we also consider a spatially dependent parameter $\beta(z)$ to modify the velocity.
It is given by 
$\beta(z)= 0.5[1-\tanh(0.1(z-50/l_0))]$,
where $l_0=1$\,Mm is the unit of length.
After the relaxation,
the unphysical velocity is suppressed to less than 2\,\kms.
Then the relaxed state is converted to 3D Cartesian coordinates $(x,y,z)$ by rotating the axisymmetric equilibrium in 2D.
Note the upper velocity layer $v'(x,z)=\beta(z)v_i(x,z)$ is still working until the end of the simulation to absorb the upward velocity when $z>50$\,Mm. 
The initial temperature and density profiles at the flux tube axis of the 3D simulation are shown in Figure~\ref{fig_ini_prof}.


To mimic the above-mentioned jet-like features,
we employ an initial velocity perturbation 
\begin{eqnarray}
  v_z(x,y,z;t=0)= v_0\exp\left(-\displaystyle\frac{x^2}{2\sigma^2_{x}}\right)
             \exp\left(-\displaystyle\frac{y^2}{2\sigma^2_{y}}\right)
             \exp\left(-\displaystyle\frac{z^2}{2\sigma^2_{z}}\right),
\label{eq_vz}
\end{eqnarray}
where $v_0=50$\,\kms, $\sigma_x=\sigma_y=\sigma_z=l_0/\sqrt{2}$.

The boundary conditions are specified as follows.
In the 2D case,
an axisymmetric boundary is employed at $r=0$,
while an outflow condition is considered at $r=100$\,Mm.
At the bottom of the flux tube, namely $z=0$,
the density and pressure are extrapolated from the hydrostatic equilibrium. As described in \cite{2023ApJ...949L...1G},
the magnetic field is extrapolated following the zero normal gradient condition \citep[see also][]{2019A&A...623A..53K}.
The vertical component of the velocity $v_z$
is set to be zero, 
while the transverse component $v_r$ is zero-gradient. 
At $z=100$\,Mm,
an outflow boundary condition is considered.
In fact,
due to the velocity absorption layer determined by $\beta(z)$,
the whole region of $z>50$\,Mm can be considered to be a no-inflow upper boundary.
In the 3D simulation, 
all the lateral boundaries are set to be outflow. 
In the $z$-direction,
all the boundary conditions are the same as in the 2D case,
except the transverse velocity $v_x$ and $v_y$ at 
$z = 0$, which is described by a continuous velocity driver
\citep[e.g.,][]{2010ApJ...711..990P,2019A&A...623A..53K,2023ApJ...949L...1G}.
It is given by
\begin{eqnarray}
  \vec{v_r}(x,y,z;t)= \vec{v_{\rm e}}+\displaystyle\frac{1}{2}\left(\vec{v_{\rm i}}-\vec{v_{\rm e}}\right)\{1-\tanh\left[b(r-1)\right]\}\cos\left(\frac{2\pi}{P}t\right),
\label{eq_v_driver}
\end{eqnarray}
where 
\begin{eqnarray}
\vec{v_{\rm i}}=v_0\hat{x},
\vec{v_{\rm e}}=v_0\left[\left(\frac{{x'}^2-y^2}{r^4}\right)\hat{x}
               +\left(\frac{2x'y}{r^4}\right)\hat{y}\right],
\label{eq_v_ie}
\end{eqnarray}
and 
\begin{eqnarray}
x'=x-\left(\frac{v_0P}{2\pi}\right)\sin\left(\frac{2\pi}{P}t\right),
r = \sqrt{{x'}^2+y^2}.
\label{eq_xper}
\end{eqnarray}
In the current model, we choose $v_0=20$\,\kms, $ P=100$\,s, and parameter $b=10$, which corresponds to the inhomogeneous layer thickness of the temperature profile, as given in Eq. (2) in \cite{2023ApJ...949L...1G}.
These driver parameters are inspired by the sinusoidal fit to the kink motions in observations (see Figure~\ref{across_jet}).

We solve the 3D ideal MHD equations by the PLUTO code \citep{2007ApJS..170..228M}.
A parabolic reconstruction method is used.
We employ the Roe-Riemann solver to compute the numerical fluxes.
A second-order characteristic tracing method is used for time marching. 
The hyperbolic divergence cleaning method is adopted to ensure that the magnetic field is divergence-free.
Anisotropic thermal conduction is also included in our simulations, 
and more details can be found in \cite{2023ApJ...949L...1G}.
The computational domain is [0,6]\,Mm $\times$ [0,100]\,Mm in the 2D run.
In the $r$-direction,
we consider 128 uniform cells,
while in the $z$-direction,
a uniform grid of 1024 cell points is adopted.
In the 3D run,
the computation domain is [-6,6]\,Mm $\times$ [-6,6]\,Mm $\times$ [0,100]\,Mm.
We consider 512 uniform grid points in the $z$-direction.
256 uniformly spaced cells are used in the $x$-and $y$-directions, respectively.

\subsection{Numerical Results and Comparison with Observations}\label{sec:results}

\begin{figure*}[!ht]
\centering
\includegraphics[width=1.0\textwidth]{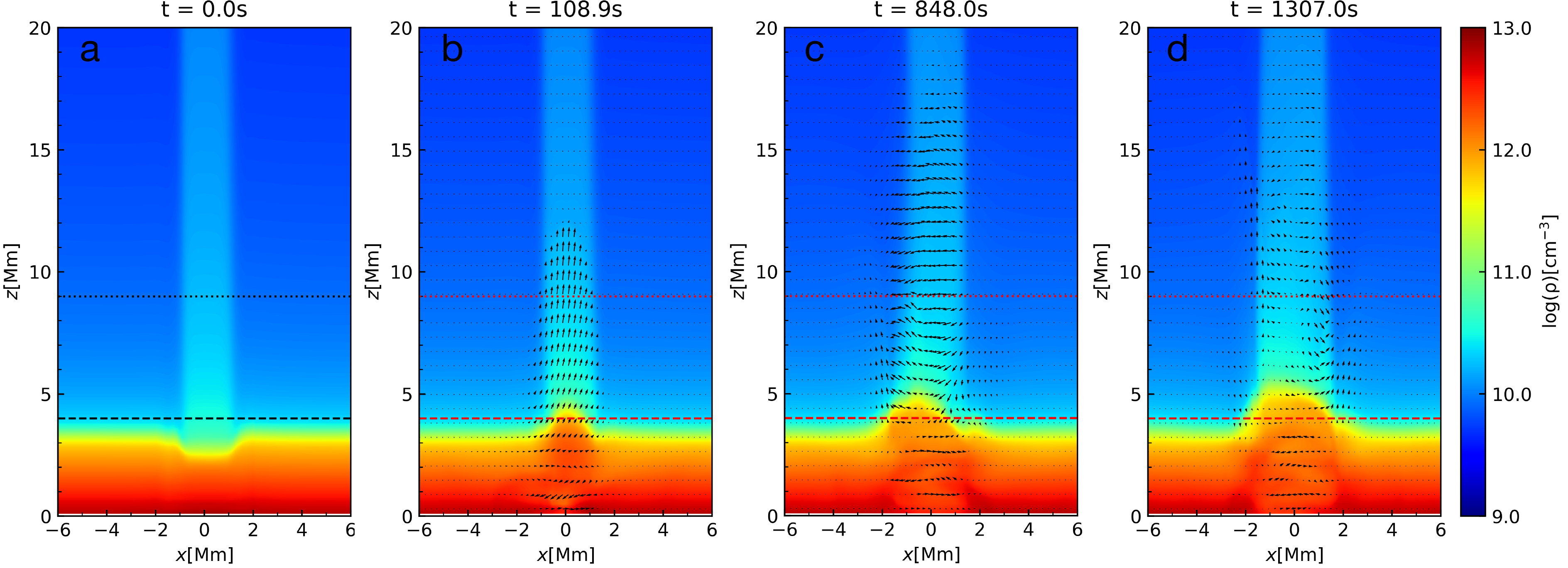}
\caption{Snapshots of density evolution at $y=0$ surface. 
Dashed (dotted) red lines mark the height $z=4$\,Mm ($z=9$\,Mm). Velocity vectors are shown by black arrows. 
An animation of the density evolution is also available. 
The animation proceeds from $t = 0$ to 1540\,s, with a total duration of 1540 seconds.}
\label{fig_snapshot}
\end{figure*}

The vertical motion excites a jet-like structure that resembles the observed chromospheric dark jet shown in Figure~\ref{fig_snapshot}. 
A similar configuration was previously investigated in 2D simulations by \citet{2024SoPh..299...53Z}. 
From the animation, 
both upward and downward motions can be identified, 
consistent with observed behaviour in chromospheric jets. 
As in the 2D case, we see rising material forming a spicule-like structure.
Downward and upward velocities near the boundary of the spicule-like structure are still evident, as shown in Figure~\ref{fig_snapshot}(c) and (d).
However, 
the formation of fine density structures due to the pre-existing plume is not clear in the current 3D model.
This may result from deformation of the transverse density structure in the chromosphere by the current 3D driver, 
which suppresses vertical pressure gradients. 
Consequently, standing waves, such as those reported by \citet{2011ApJ...736L..24O} in spicules, 
are not clearly manifested in the current model due to the deformed density structure.

\begin{figure*}[!ht]
\centering
\includegraphics[width=\textwidth]{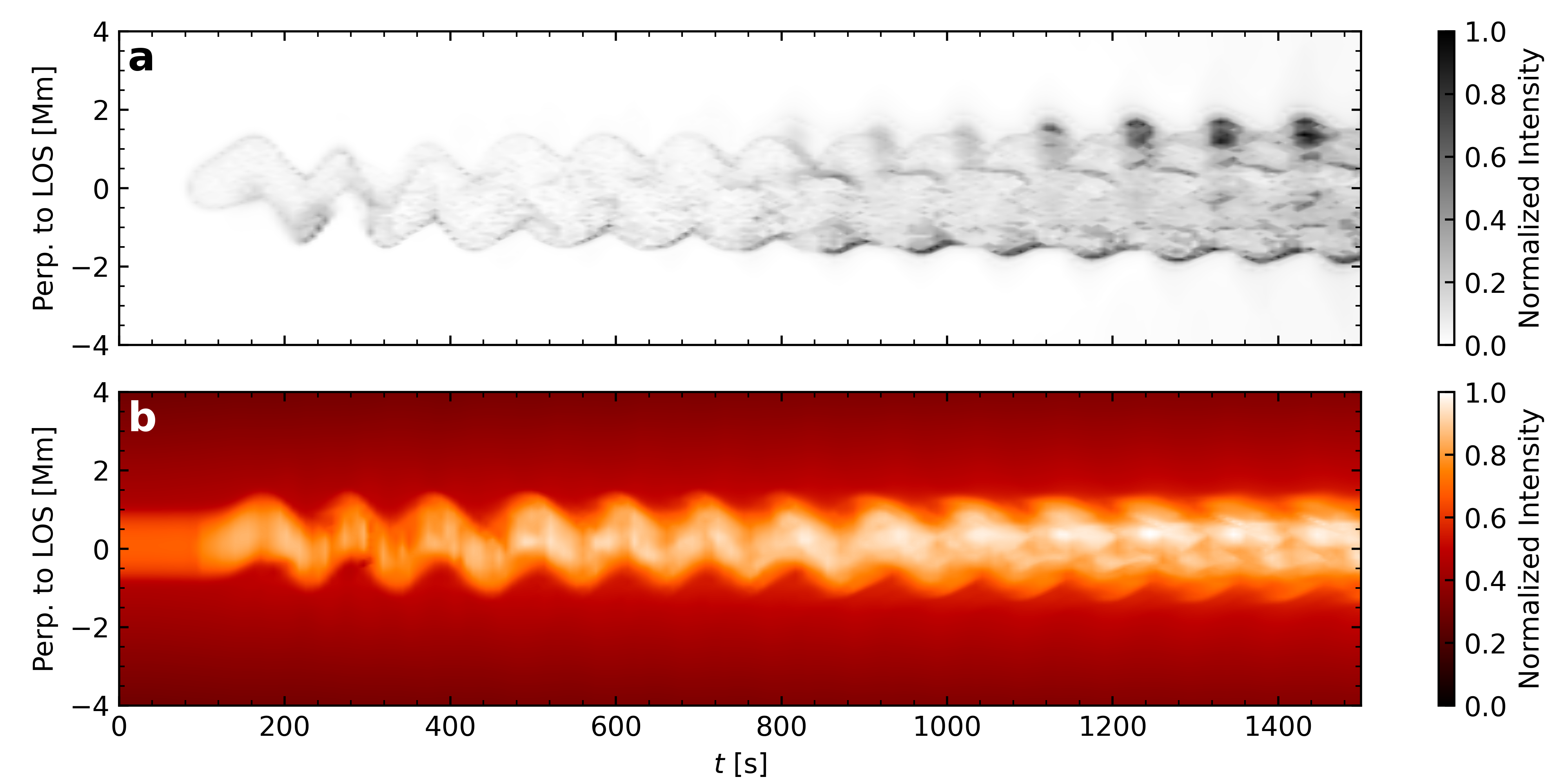}
\caption{
Time-distance maps of the forward-modelled intensity from the numerical simulation in (a) the 304~\AA\ line at $z=4$~Mm and (b) the 174~\AA\ line at $z=9$~Mm, for a LOS angle of $45^\circ$.
 }
\label{fig_td_map}
\end{figure*}

\begin{figure*}[!ht]
\centering
\includegraphics[width=0.8\textwidth]{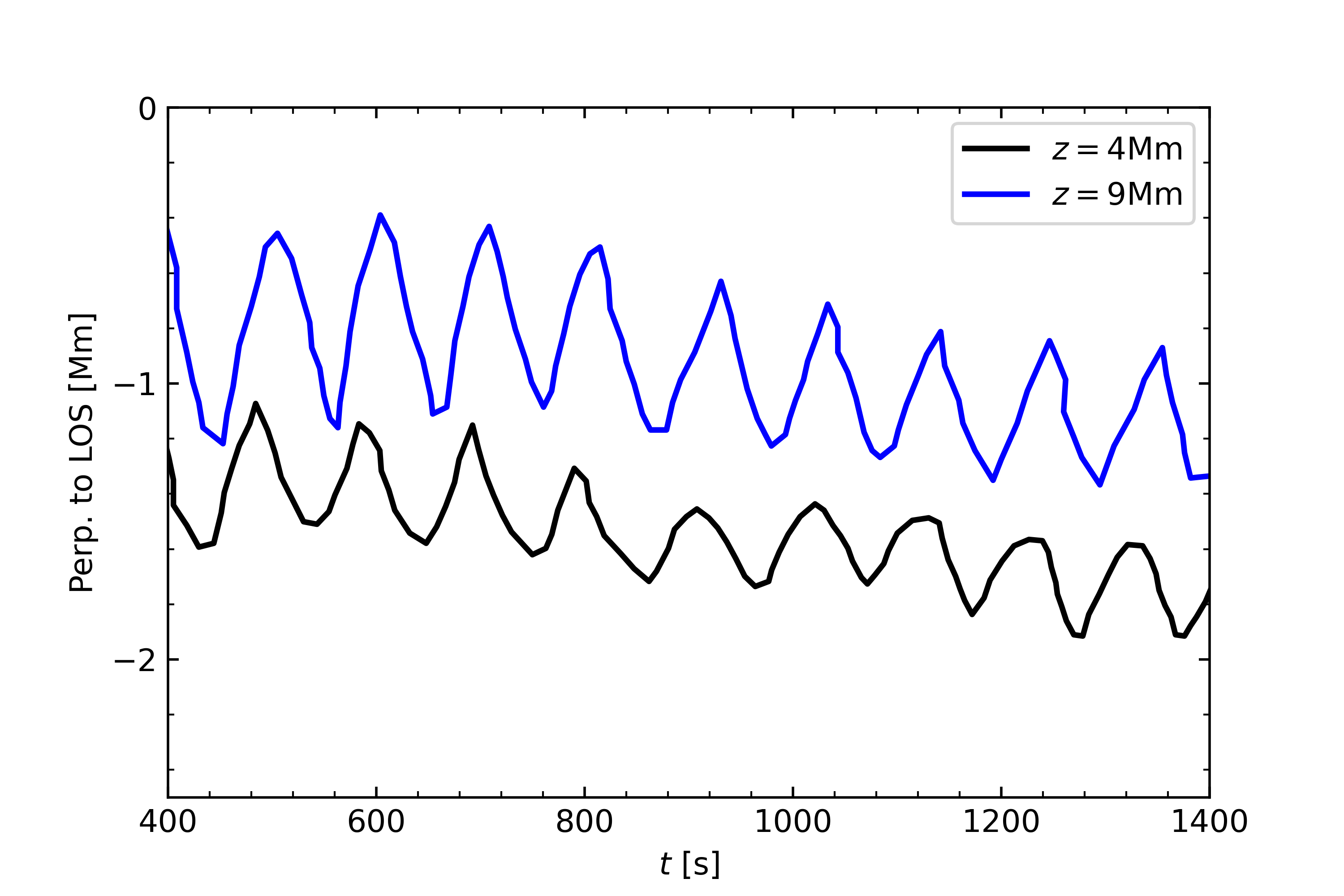}
\caption{Oscillation profiles of the edge of the density shown in Figure~\ref{fig_td_map} from $t=400$\,s to 1400\,s. 
Black (blue) line represents the oscillation profile at $z=4$\,Mm ($z=9$\,Mm). }
\label{fig_profiles}
\end{figure*}

To facilitate comparison with the observations presented in Section~\ref{sec:data},
we also obtained forward models using the FoMo code
\citep{2016FrASS...3....4V}. 
We selected the 304~\AA~and 174~\AA~emission lines, 
which are sensitive to chromospheric and coronal temperatures, respectively.
Figure~\ref{fig_td_map} shows the time-distance maps of the intensity at heights of $z=4$~Mm and $z=9$~Mm, 
with the LOS angle of $\pi/4$.
We observe transverse motions of both the chromospheric jet and the coronal plumelet, 
which result from the transverse velocity driver at the bottom.
In both panels, 
the intensity appears to increase during the second half of the time interval, especially for $t>800$~s.
We attribute this enhancement to local heating induced by wave energy dissipation 
\citep[see e.g.,][for a recent review]{2017A&A...604A.130K,2019ApJ...870...55G,2020SSRv..216..140V}.
In the present work, 
however, 
the observational resolution is not sufficient to provide further constraints on the heating process.
Nonetheless, 
we can still compare the oscillation period and amplitude of the forward models with observations.
In Figure~\ref{fig_profiles},
we plot the edges of the intensity shown in Figure~\ref{fig_td_map} over the time interval from $t=400$~s to $1400$~s.
We find that the oscillation periods of the intensity profiles at $z=4$~Mm and $z=9$~Mm are both consistent with that of the bottom driver.
This indicates that the transverse oscillation of the plumelet is essentially an upward-propagating wave originating from the chromosphere.
We also compare the oscillation amplitudes in our simulations with those measured from observations.
As shown in Figure~\ref{fig_profiles}, 
the amplitude of the chromospheric jet is approximately 170~km, 
while that of the plumelet is about 270~km.
These values are compatible with those measured at the corresponding heights in the observations (see Figure~\ref{tab_sin}), 
and the relative amplitude, i.e., larger amplitude in the corona than in the chromosphere, is also consistent with the observations.
Because the density and temperature distributions are prescribed in our simulations,
this agreement provides a potential approach to constraining density and temperature values and distributions that are not straightforwardly measurable from observations.

Note that in the current model,
the plume structure is pre-existing as a density-enhanced structure. 
While the chromospheric jet is not pre-existing, 
it is excited by the vertical velocity driver at the bottom. 
As aforementioned, 
this is inspired by observations that the chromospheric jet and the plumelet share the same magnetic skeleton,
as illustrated in Figure~\ref{fig_fov}(b2).
Although the initial density ratio between the internal and external flux tube at the bottom ($z = 0$) is set to 3, 
as shown in Figure~\ref{fig_ini_prof}(a), 
this ratio approaches unity after relaxation (see Figure~\ref{fig_ini_prof}(b) and Figure~\ref{fig_snapshot}(a)).
As a result, 
the transverse density in the chromosphere becomes nearly uniform, 
confirming that the chromospheric jet shown in 
Figure~\ref{fig_snapshot} is indeed excited by the vertical velocity perturbation.

In this section,
a numerical simulation is conducted,
inspired by previous observations,
a transverse wave is excited in a pressure-driven chromospheric jet, 
which subsequently propagates upward into the associated coronal  plumelet.
The simulation successfully reproduces key features of the observations, demonstrating that kink waves can carry energy from the lower solar atmosphere into the corona.

\section{Summary}
\label{sec:con}
In this work, we analyzed propagating waves in chromospheric jets and their direct correspondence to coronal plumelets observed by 174\,\AA~HRI$_{EUI}$ aboard the Solar Oribter.
The observations reveal that the chromospheric darkjet-like structure exhibits two directional motions: upward motion and transverse motion during its lifetime. 
The dark jet that we are concerned with has an upward speed of 26\,\kms and a lifetime of 230\,s. 
For the transverse motion, 
it shows a period of about 95\,s and a displacement of about 193\,km.
The plumelets, which are substructures within the plumes and in conjunction with dark jets, 
also exhibit transverse motions with a displacement of approximately 315\,km and an oscillation period of about 99\,s. 
Given the similar oscillation period of the dark jet and the plumelet,
we state that the observed oscillation is a transverse propagating wave originating from the chromosphere.  
The larger oscillation displacement of the plumelet is attributed to the density stratification within the open magnetic structures. 
Furthermore, we estimated that the propagating speed of the transverse wave is about 91\,\kms, which is close to the Alfv\'en speed in the chromosphere \citep[][]{2007Sci...318.1574D}. 
We thus believe that the transverse wave observed is an upward propagating kink wave.  
To prove the above statement, we conducted a 3D MHD simulation.
Vertical and transverse displacements were launched into a stratified magnetic flux tube.
The main dynamic properties of the chromospheric jet and the plumelet have been reproduced.

The current study confirms the propagation of high frequency kink waves from the chromosphere into the corona through direct observations. 
Although transverse waves have been widely reported in chromospheric structures and are known to be ubiquitous in coronal plumes
~\citep[e.g.,][]{2007Sci...317.1192T,2007Sci...318.1574D,2009ApJ...705L.217H,2014ApJ...790L...2T,2015NatCo...6.7813M},
the periods measured in previous studies of coronal plumes are generally longer than those reported here, 
largely owing to lower temporal resolution. 
Recently, high resolution observations from the DKIST have confirmed 
the presence of higher frequency (above 10mHz) propagating Alfv\'enic waves~\citep[e.g.,][]{,2025ApJ...982..104M,2025ApJ...991...97H}.
In fact, 
transverse waves are thought to have cutoff frequencies across the transition region \citep[e.g.,][]{2017AJ....154..141L,2023A&A...672A.105P}, 
which can limit their ability to transport energy upward. 
However, kink waves with higher frequency can reach the corona by tunneling through the transition region. 
In this context, our current finding of kink propagating waves with a period of 100s provides further evidence 
that transverse waves may contribute to energy transfer from the lower atmosphere to the upper corona.

In addition, 
this study offers a possible mechanism for the excitation of kink waves in the solar corona. 
Traditionally, 
coronal waves are believed to originate from photospheric motions \citep[e.g.,][]{2014ApJ...784...29M} or from magnetic reconnection associated with solar eruptive activities \citep[e.g.,][]{2025ApJ...982L..25Y}. 
Here, we suggest that high frequency coronal kink waves may also originate in the chromosphere, 
although the mechanism for their excitation in the chromosphere remains unclear. 
Nevertheless, 
the current results pave the way for future studies aimed at detecting chromospheric dynamics and clarifying their dynamical and energetic connections to the upper corona.

\nolinenumbers
\begin{acknowledgments}

This work is supported by the National Natural Science Foundation of China (No. 42230203,42174201) and the Shandong Province Natural Science Foundation (ZR2023QD14).
MG acknowledges the support from the National Natural Science Foundation of China (12203030), the QILU Young Scholars Program of Shandong University, the Taishan Scholars Program (tsqn202408051) and the Shandong Provincial Natural Science Foundation for Excellent Young Scientists Program, Overseas (2025HWYQ-019).
TVD was supported by the C1 grant TRACEspace of Internal Funds KU Leuven, and a Senior Research Project (G088021N) of the FWO Vlaanderen. Furthermore, TVD received financial support from the Flemish Government under the long-term structural Methusalem funding program, project SOUL: Stellar evolution in full glory, grant METH/24/012 at KU Leuven. The research that led to these results was subsidised by the Belgian Federal Science Policy Office through the contract B2/223/P1/CLOSE-UP. It is also part of the DynaSun project and has thus received funding under the Horizon Europe programme of the European Union under grant agreement (no. 101131534). Views and opinions expressed are however those of the author(s) only and do not necessarily reflect those of the European Union and therefore the European Union cannot be held responsible for them.

\end{acknowledgments}

\clearpage
\bibliographystyle{aasjournal}
\bibliography{./darkjet_plume}


\end{document}